# An Architectural Approach for Decoding and Distributing Functions in FPUs in a Functional Processor System

[1] T.R.Gopalakrishnan Nair, [2]R. Selva rani, [3]Krutthika H.K
[1]Director, Research & Industry Incubation Center, Senior Member IEEE trgnair@ieee.org
[2]Associate Professor, Research-CSE, selvss@yahoo.com
[3]M.Tech II Year, krutthika.hk85@gmail.com
[1,2,3]Dayananda Sagar Institutions, Bangalore, India

**Abstract – The main goal of this research is to develop the concepts of a revolutionary processor system called Functional Processor System. The fairly novel work carried out in this proposal concentrates on decoding of function pipelines and distributing it in FPUs as a part of scheduling approach. As the functional programs are super-level programs that entails requirements only at functional level, decoding of functions and distribution of functions in the heterogeneous functional processor units are a challenge. We explored the possibilities of segregation of the functions from the application program and distributing the functions on the relevant FPUs by using address mapping techniques. Here we pursue the perception of feeding the functions into the processor farm rather than the processor fetching the instructions or functions and executing it. This work is carried out at theoretical levels and it requires a long way to go in the realization of this work in hardware perhaps with a large industrial team with a pragmatic time frame.**

***Index Terms –*** *Functions scheduling**,** Functional Processor Unit (FPU), Fine Decoding**, First**-In First-Out, Funpiler.*

## 1. INTRODUCTION

The most recent advances in microprocessor design for desktop computers involve putting multiple processors on a single computer chip. These multicore designs are on its way to completely replace the traditional single core designs that have been the foundation of desktop computers. The race to control the market share in this new area has forced each computer chip manufacturer to push the envelope of the number of cores that can be economically placed on a single chip. All of this competition places more computing power in the hands of the consumer than ever before. The goal of this work is to design a novel, pervasively parallel architecture called "Functional Processor Architecture" which is an innovative architecture at research level.

The software industry has witnessed a massive growth since its transformation from a pristine sector, catering only to scientific research. This exponential growth has posed a lot of challenges related to programming aspects and the cost of software packages. The available multicore processors and programming models, though partially useful, have deemed insufficient for a revolutionary growth scenario. The need of the hour is a comprehensive and innovative design approach in architecture [1].

The perspectives and the motivation for this work have come from the fields of object-oriented design. Usually, human thoughts and programming models are forced to structure itself with the architecture of processors and to adjust the freedom of programming in connivance with the processors. Here the work is carried on the "Functional Processor Architecture" as an innovative architecture at research level. This paper will briefly discuss the influence of novel architectural design of object oriented paradigm. The most influencing architectural properties for this are identified and an effective scheduling algorithm using First-in First-out (FIFO) is adopted. A Graph theoretical approach is used in analyzing the Function dependency.

The main goal of this research is to generate a viable functional processor system whereby a program is represented as a sequence of higher level functions only and get executed on multiple functional processor units. It also intents to demonstrate a complete deviation from a conventional fetching system to a forced feed system to processors.

## 2. CHALLENGES

Currently, the performance of Multicore processor depends on the problem being solved and the algorithms used as well as their implementation in conventional software models. The mismatch in total power of number of processors and resultant speed is mainly because of the constraints in the programming language and the architecture of the processors [4]. The functional processor architecture tries to answer these challenges by providing the multiprocessors with a significant number of dedicated programmable cores targeting a broad set of workloads, including intensive multimedia and scientific processing functions available at FPUS.

## 3. FUNCTIONAL PROCESSOR ARCHITECTURE OVERVIEW

The Functional Processor Architecture (FPA) is expected to addresses the needs of applications as they embrace multiprocessing. Rather than merely replicating a core multiple times on a chip, the FPA's heterogeneous architecture offers a mix of execution elements optimized with an array of functions which can be used as complex instructions. Applications get executed on this system by partitioning the application and processing each component on the most appropriate execution element. While supporting different execution elements, the architecture also ensures efficient sharing by providing a common systemic view of address system, data types and system functions across the heterogeneous execution elements.

Any program contains an ordered, designed set of functions and is given as an input to the functional decoder. The functional decoder identifies the functions from the program and divides the Functions, say $Fn_1$ ,$Fn_2$…..$Fn_n$. These functions are then pushed into a fine decoding block. Here the functions are analyzed, classified and is finally fed into the suitable Functional Processing Unit (FPU) along with developing a graph theory based placement and scheduling approach.

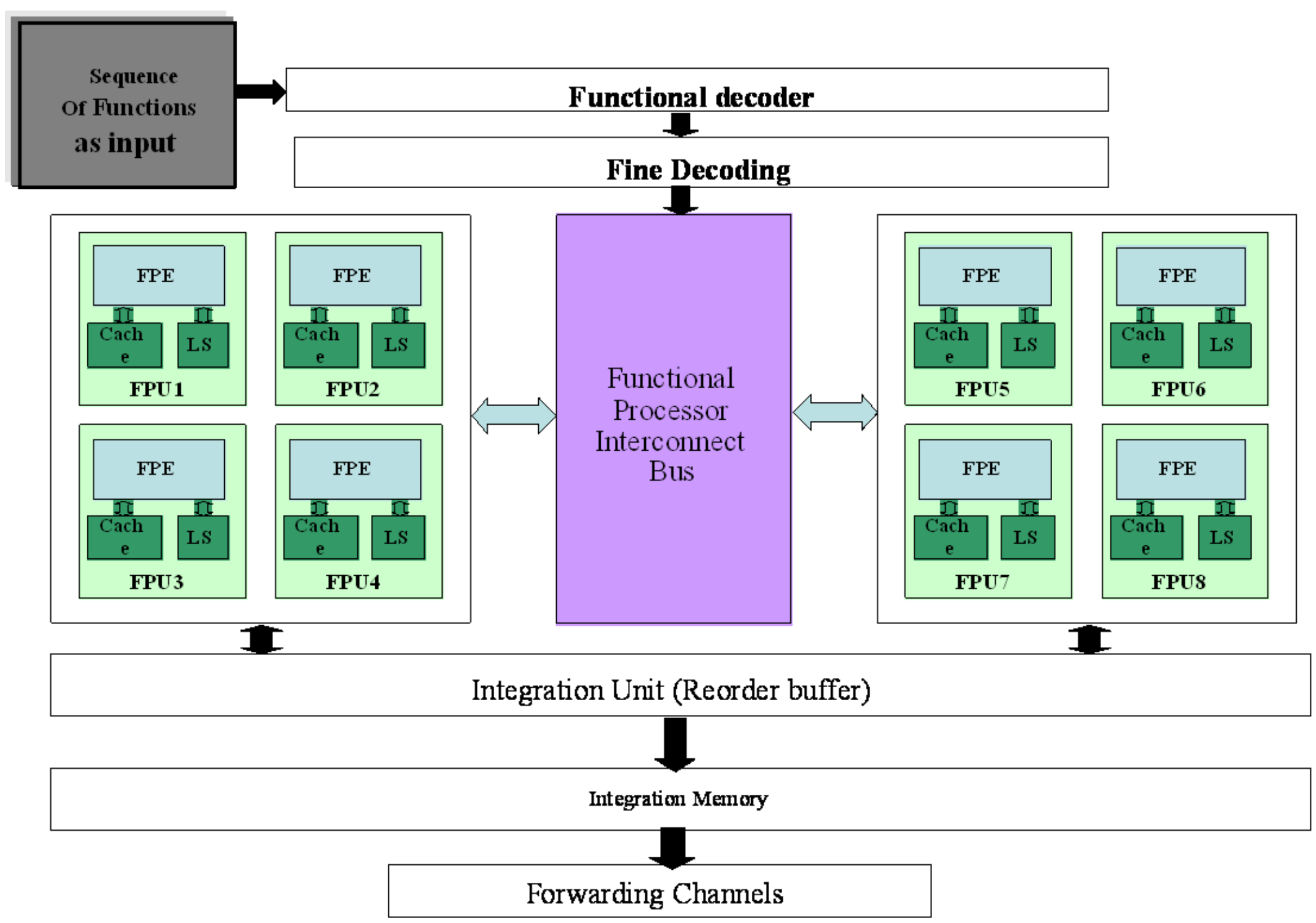

Figure 1. Functional Processor Architecture
FPU: Functional Processor Unit, FPE: Functional Processor Element, LS: Local store

Here these cores (FPUs) are with dedicated properties; say FPU1 may be a graphical FU, FPU2- arithmetic FU, FPU3- a DSP FU, FPU4- a string processor etc.. According to the requirements of the application, the functions references of main program are fed into appropriate functional processor unit (FPUs) with the help of the functional processor interconnect bus for detailed processing. The FPU executes these already defined functions accordingly. The corresponding –related- functional programs are stored in local store. It is like a library of functions.

Whenever the executing function wants any related function, then the FPU automatically fetches from the local store through a ultra high speed bus. In the integration part the sequencing of functions play a significant role. The output from the integration block is stored in the memory unit for the further use.

The simulation was done using C++ programming language. The "Funpiler" is the function compiler in Fine decoding block assigns addresses to the functions in the application program i.e., function ID (FID). If the function requires DSP processor then FID D1, FID D2 etc. are assigned to the functions. If the function requires arithmetic processor then FID A1, FID A2 etc. are assigned to the functions. If the addresses of the functions in the functional decoder maps to the arithmetic processor or DSP processor then the functions are fed and execution takes place there.

We have adopted First in First out (FIFO) scheduling algorithm also called First-Come First-Served. Here, whichever functions come first to the FPU that function dominates till its end of execution. The functions either can be dependent or independent. There is no special mechanism for the functions to communicate with each other while being executed. Dependency is created in the sense that the output from the $Fn_1$ is given as input to the $Fn_q$ etc. and also to the other functions if needed and vice versa.

## 4. FUNCTION EXECUTION IN FPU COLONY

In order to extort the maximum throughput from the heterogeneous function processor colony, it is important to keep all the FPUs balanced, avoiding bottlenecks whenever possible. If we design a high-performance processor capable of executing ' n' FPU operations at once, it is also important to ensure that we can feed the functions into FPU stage and retire those functions without stalling the pipeline. This means feeding and decoding at least optimal set of functions per cycle, to keep the FPU stage busy and writing results at a faster rate.

In this paper, we have proposed function fed mechanism by eliminating the program counter approach, which is used in conventional processor architectures. In the function fed mechanism, the memory latency is low because we are concentrating only on functions in utltra memory instead of instructions sought from low speed ones and also the processor time is reduced substantially by allowing the processor to execute an array of functions in FPUs. Where as, in an instruction fetching mechanism with the style of executing only one instruction at a time, the obstacle faced by the fetch mechanism through the memory latency is large. The time it takes involves the time to read an instruction from memory together with the time taken to execute the instruction. If the memory latency is large, it quickly becomes the major component of the processor time. Current model employs FPUs of conventional type processors at nodes but it is decided to use Push Instruction Format ( PIF) architecture later, similar to that we employed at FPU farm approach.

The fine decoding feeds the functions into the corresponding dedicated cores and also maximizing system throughput and ensuring fairness among the running of functions in the system. The throughput of an application can be increased with parallel multiple functions. With one function, an I/O would halt the entire process, with multiple functions, as one function waits for an I/O request, the application continues to execute. The computational speed of the functional processors may strengthen by adding FPUs for Input Output process too.

Figure-2 is the flowchart for functional decoder. An application contains an array of functions and these functions are given as input to the functional decoder and are stored in the memory. Here read operation takes place and the functional decoder identifies the functions from the process. If it finds the functions, then the functions are separated from the process and these functions are stored in separate modules.

Figure-3 is the flow chart for the Fine Decoding block. After the splitting of functions from the process, the functions are fed into the appropriate FPUs. Here in the fine decoding stage, the "Funpiler" is the function compiler that analyzes the functions. If the function requires the arithmetic processor then the FIDs are assigned say FID A1, FID A2 etc. to the functions.

### 4.1 Function anatomy

Figure-4 shows the architecture of a process which contains multiple functions. Both have context and attributes that makes the process unique from other processes in the system and attributes that makes a function unique from its peer functions. The process global variables are located in the data segment. The context for functions Fn1 and Fn2 has function ids, states and priority. This will be appropriately made use of in creating temporal execution chart and connectivity.

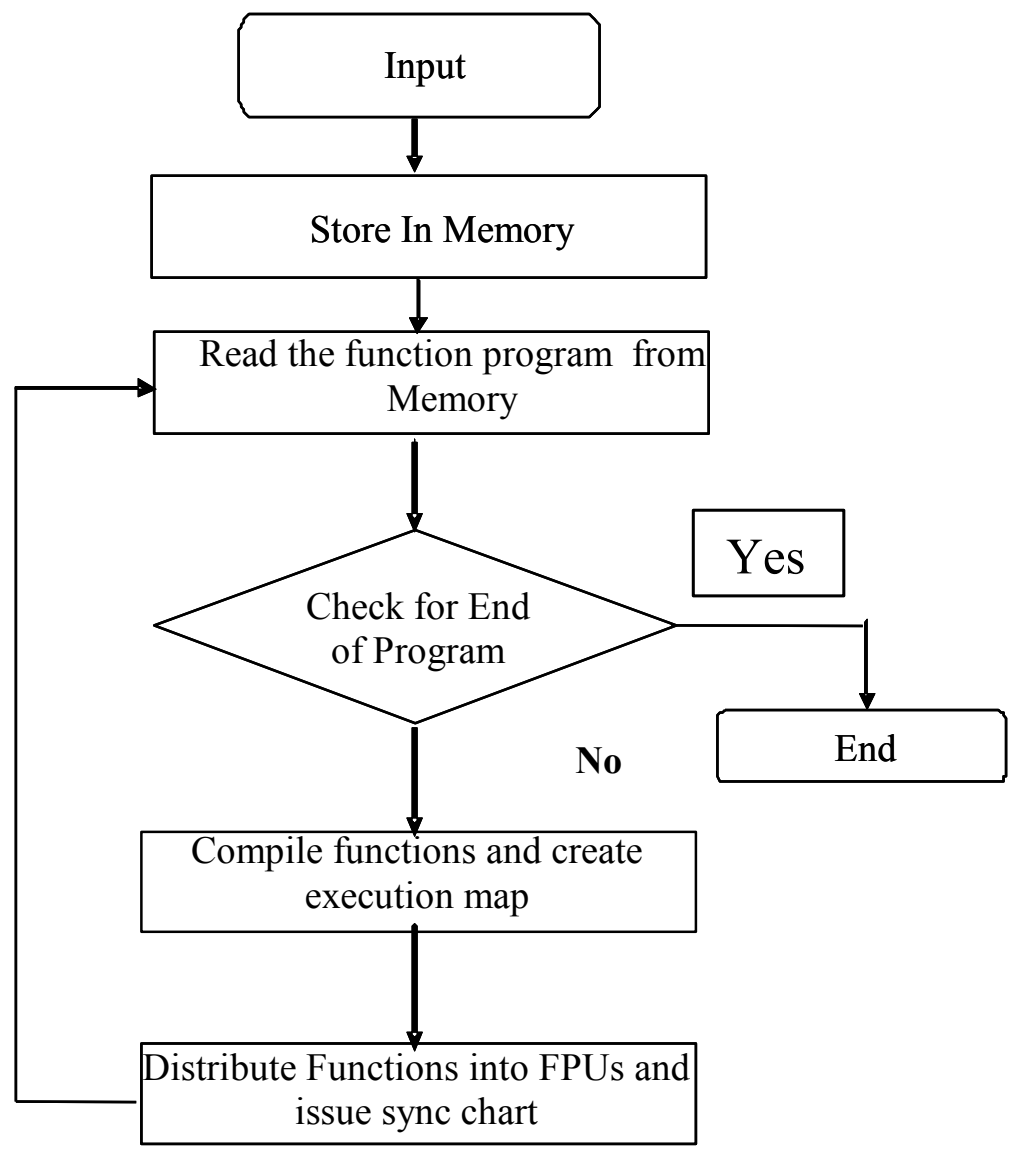

Figure 2. Flow chart for Functional decoder

## 5. FIFO SCHEDULING POLICY

Figure-5 depicts the scheduling scheme adopted in the functional processor. With a FIFO scheduling policy, functions are assigned to the processor according to the arrival time in the queue. A function assigned to a FPU dominates the processor until it completes execution. This scheduling policy can be used for applications where a set of functions need to complete as soon as possible. When a sleeping function becomes runnable, the function is placed at the end of its priority queue. A function can make a system call and give up the processor to another process with the same priority level. The function is then placed at the end of its priority queue.

### 5.1 Scheduling of Functions

The fine decoding contains several operations. The scheduler must determine which function should be assigned to what function processor. The scheduler maintains data structures that allow it to schedule the functions in an efficient manner. Each function is given a priority class and placed in a fine decoding (priority queue) with other executable functions with the same priority class. There are multiple priority queues, each representing a different priority class used by the system. These priority queues are stratified and placed in a dispatch array called the multilevel functional priority queue. Figure-5 depicts the multilevel functional priority queue. Each element in the array points to a priority queue.

Priorities can be dynamic or static. Once a static priority of a function is set, it cannot be changed. Dynamic priorities can be changed. Functions with the highest priority can monopolize the use of the processor. If the priority of a function is dynamic, the initial priority can be adjusted to a more appropriate value.

The function placed in a priority queue has a higher priority. A process monopolizing the processor can also be given a lower priority, or other functions can be given a higher priority than that process has. When you are assigning priority to a user functions, consider what the function does most of its time. Some functions are FPU - intensive. FPU intensive functions use the processor for the whole time slice. Some functions spend most of its time waiting for I/O or some other event to occur. When such a function is ready to use the processor, it should be given the processor immediately so it can make its next request for I/O. Functions that are interactive may require high priority to assure good response time. System functions have a higher priority than user functions. The functions are placed in a priority queue according to a scheduling policy.

## 6. SUMMARY

The function processor system was designed to bring a new level of computing capability to applications. The design challenge for the functional processor was to manage the triple constraint of performance, operational philosophy, and the simulation of basic units in an aggressive schedule. The sample functional processor contains eight processor cores that can have eight parallel function units active at any one time. The design and methodology rely on the hierarchy and sequence of executing functions on different FPUs synchronously.

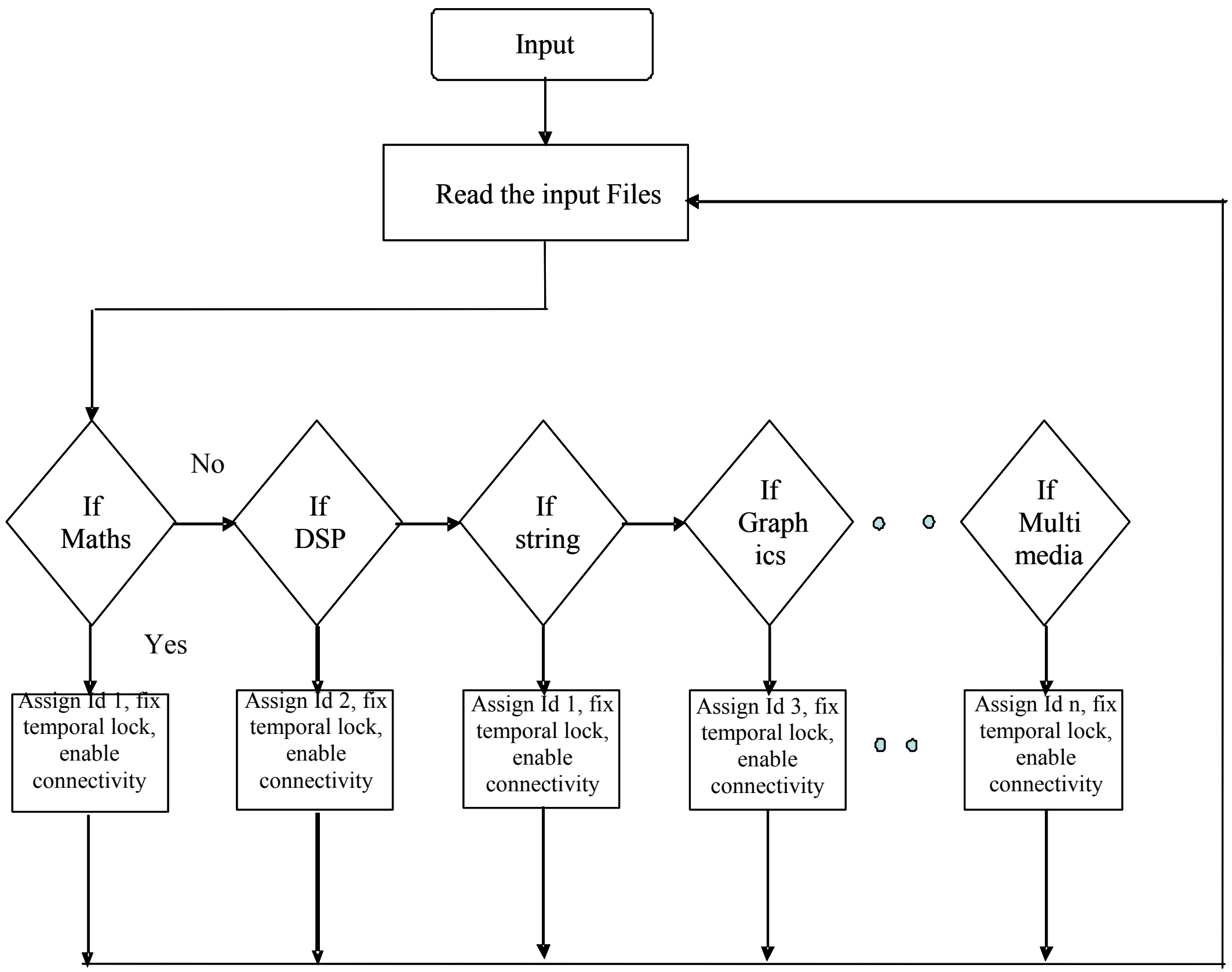

Figure-3 Flow Chart for Fine Decoding.

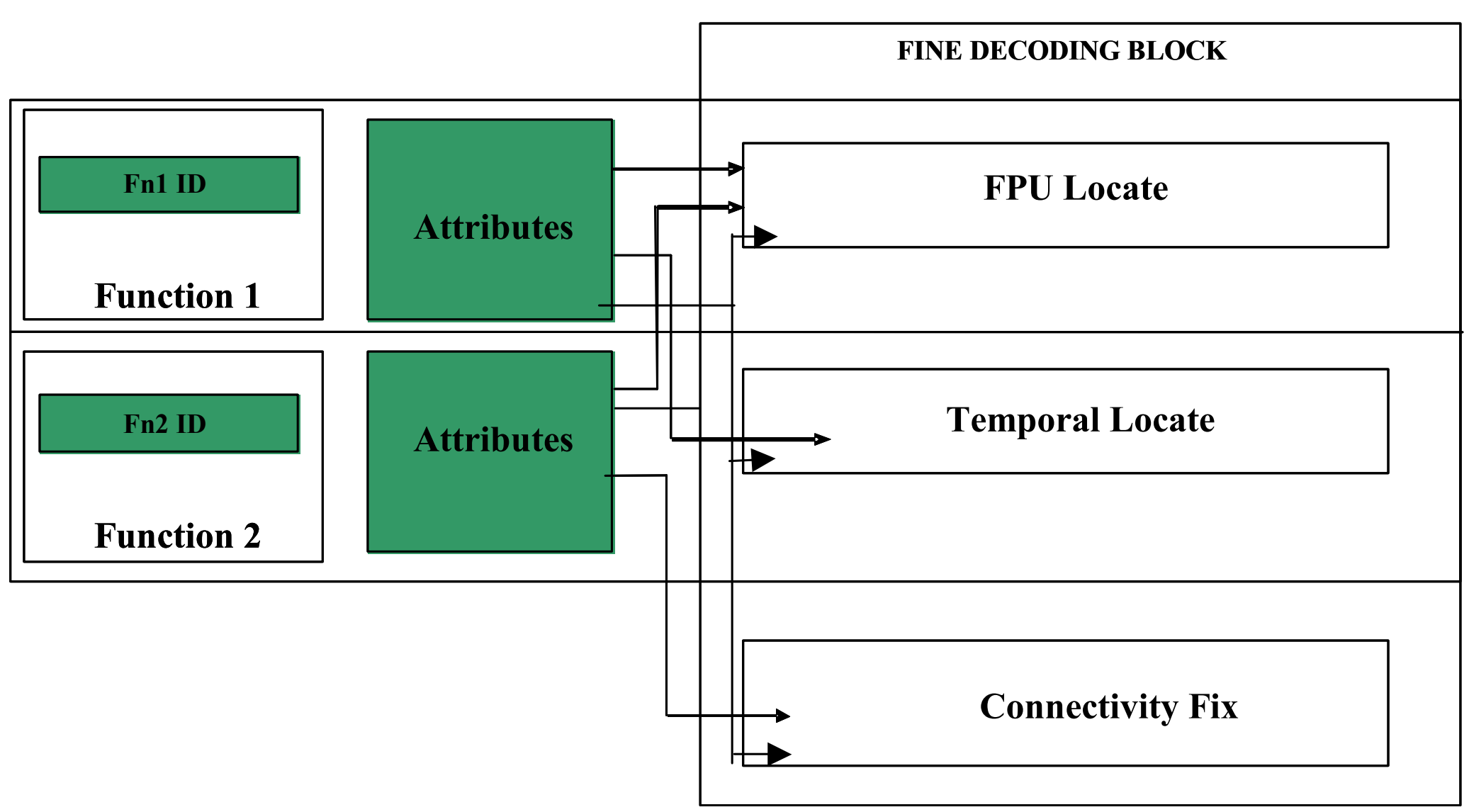

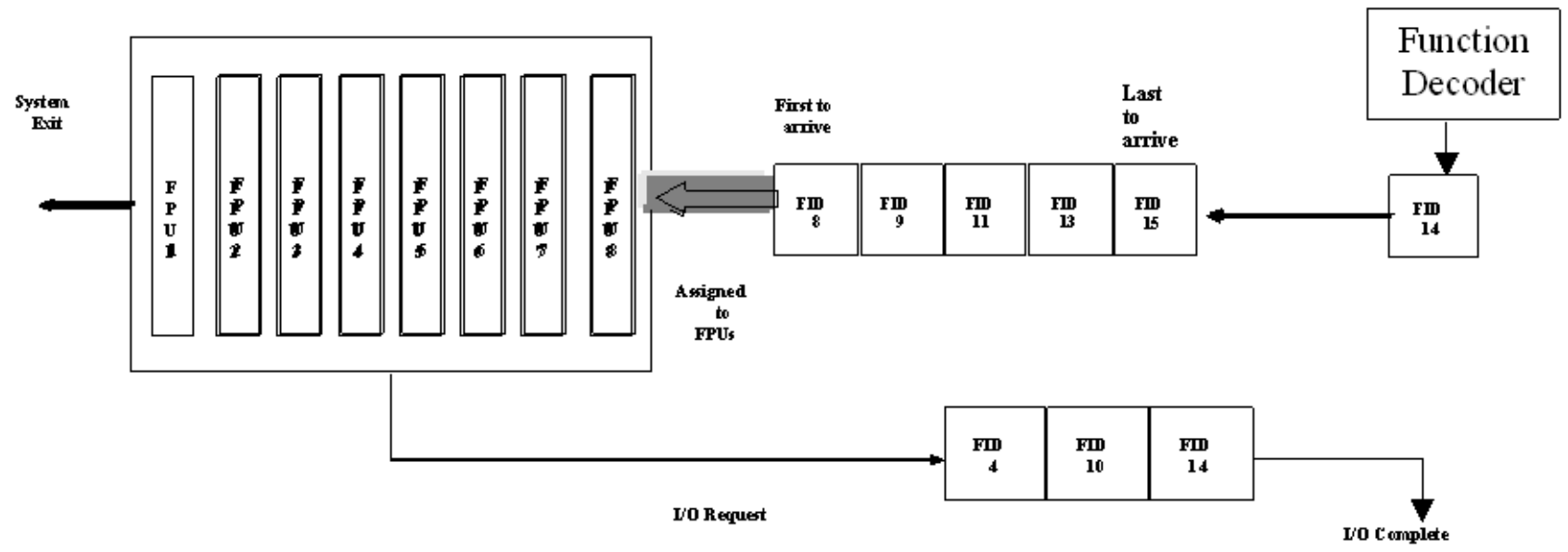

Figure 5. Exemplifies FIFO in operation

## REFERENCES

[1] Ibrahim Hur, Calvin Lin "Adaptive history-based memory schedulers For Modern processors", Published by the IEEE Computer Society, 0272-1732/06/$20.00 © 2006 IEEE.

[2] Jonathan Owen Maurice Steinman, "Northbridge architecture of AMD's griffin microprocessor family", Published by the IEEE Computer Society. 0272-1732/08/$20.00 *G* 2008 IEEE.

[3] Pat Conway Bill Hughes, "The AMD opteron Northbridge architecture", Published by the IEEE Computer Society. 0272-1732/07/$20.00 G 2007 IEEE.

[4] Mark D. Hill and Michael R. Marty, "Amdahl's Law in the Multicore Era".

[5] Kenji Kise, Takahiro Katagiri, Hiroki Honda, and Toshitsugu Yuba,"Super Instruction-Flow Architecture for High Performance and Low Power Processors" Proceedings of the Innovative Architecture for Future Generation High-Performance Processors and Systems (IWIA'04) 1527-1366/04 , IEEE

[6] Ayose Falcón, Jared Stark, Alex Ramirez, Konrad Lai, Mateo Valero, "Better branch prediction through prophet/critic hybrids", Published by the IEEE Computer Society 0272-1732/05/ ..2005 IEEE.

[7] Erik Lindholm,John Nickolls, Stuart Oberman, John Montrym,Nvidia. Nvidia Tesla: A Unified Graphics And Computing Architecture", Published by the IEEE Computer Society, 0272-1732/08/, *G* 2008 IEEE.

[8] Kyle J. Nesbit James E. Smith, University of Wisconsin–Madison, Miquel Moreto, Polytechnic University of Catalonia Francisco J. azorla, Barcelona Supercomputing Center, Alex Ramirez,Mateo Valero, Barcelona Supercomputing Center and Polytechnic University of Catalonia, Kyle J. Nesbit, James E. Smith,University of Wisconsin–Madison, Miquel Moreto Polytechnic University of Catalonia, Francisco J. Cazorla, Barcelona Supercomputing Center. Multicore Resource Management", Published by the IEEE Computer Society 0272-1732/08/, *G* 2008 IEEE.

[9] Wen-mei Hwu, Shane Ryoo, Sain-Zee Ueng, John H. Kelm, Isaac Gelado†, Sam S. Stone, Robert E. Kidd, Sara S. Baghsorkhi, Aqeel A. Mahesri, Stephanie C. Tsao, Nacho Navarro†, Steve S. Lumetta, Matthew I. Frank, and Sanjay J. Patel, "Implicitly Parallel Programming Models for Thousand-Core Microprocessors", *DAC* 2007, June 48,2007, San Diego, California, USA

[10] David Geer,"Chip Makers Turn to Multicore Processors".Published by the IEEE Computer Society, May 2005.

[11] Pam Frost Gorder,"Multicore Processors For Science And Engineering", Co published by the IEEE CS and the AIP 1521-9615/07/ © 2007 IEEE

[12] John Bresnahan, 1, 2, 3 Rajkumar Kettimuthu,1, 2 Mike Link,1,2 Ian Foster1,2,3 "Harnessing Multicore Processors for High Speed Secure Transfer",Mathematics and Computer Science Division, Argonne National Laboratory, Argonne, IL, 1-4244-1580-2/07/$25.00 ©2007 IEEE.

## BIOGRAPHY

**T.R. Gopalakrishnan Nair** holds M.Tech. (IISc, Bangalore) and Ph.D. degree in Computer Science. He has 3 decades experience in Computer Science and Engineering through research, industry and education. He has published several papers and holds patents in multi domains. He has won the PARAM Award for technology innovation. Currently he is the Director of Research and Industry in Dayananda Sagar Institutions, Bangalore, India.

**R. Selvarani** holds MI.Tech., Ph.D. in Computer Science and Engineering (Thesis submitted) and has 18 years of experience in teaching and research. She has published several research papers in computer science and holds two patents. She has been awarded Best Teacher Award twice in various institutions. Currently she is working as a Professor, Research in Research and Industry incubation centre, Dayananda Sagar Institutions, Bangalore, India.

**Krutthika H.K**, holds M.Tech degree (Thesis submitted) in Digital communication and Networking branch. She has 1Year of experience in software industry. She is a Fellow of IEEE.